\documentclass[10pt,preprint]{revtex4-1}

\usepackage{amsmath}
\usepackage{amssymb}
\usepackage{graphicx}
\usepackage{color}

\usepackage[letterpaper]{geometry}
\usepackage[scriptsize,justification=raggedright]{caption}

\def\be{\begin{equation}}
\def\ee{\end{equation}}

\newcommand{\aleq}{\mbox{\
\raisebox{-.9ex}{$\stackrel{\textstyle<}{\sim}$}\ }}
\newcommand{\ageq}{\mbox{\
\raisebox{-.9ex}{$\stackrel{\textstyle >}{\sim}$}\ }}

\begin{document}

\title{Analytical Results for Size-Topology Correlations in 2D Disc and Cellular Packings}
\author{Matthew~P.~Miklius}
\affiliation{ Engineering Sciences and Applied Mathematics, Northwestern University, Evanston, IL, USA}

\author{Sascha~Hilgenfeldt} 
\affiliation{Mechanical Science and Engineering, University of Illinois at Urbana-Champaign, Urbana, IL, USA}

\begin{abstract}
Random tilings or packings in the plane are characterized by a size distribution of individual elements (domains) and by the statistics of neighbor relations between the domains. Most systems occurring in nature or technology have a unimodal distribution of both areas and number of neighbors. Empirically, strong correlations between these distributions have been observed and formulated as universal laws. Using only the local, correlation-free granocentric model approach with no free parameters, we construct accurate analytical descriptions for  disc crystallization, size-topology correlations, and Lema\^itre's law.
\end{abstract}

\maketitle

Materials with a cellular structure of domains on mesoscopic scales constitute a large variety of industrially important materials (e.g.\ emulsions, foams, polycrystalline metals, ferromagnets) as well as the vast majority of living tissues in multicellular organisms \cite{thompson42,lecuit07}. A description of the geometric and topological properties of the domains is crucial for the understanding of the material. Most cellular materials exhibit a degree of disorder, so that sizes, shapes, orientations, or neighbor relations of domains are determined through statistical distributions functions. 
 
\begin{figure}[b]
 \centering
\includegraphics[width=3in]{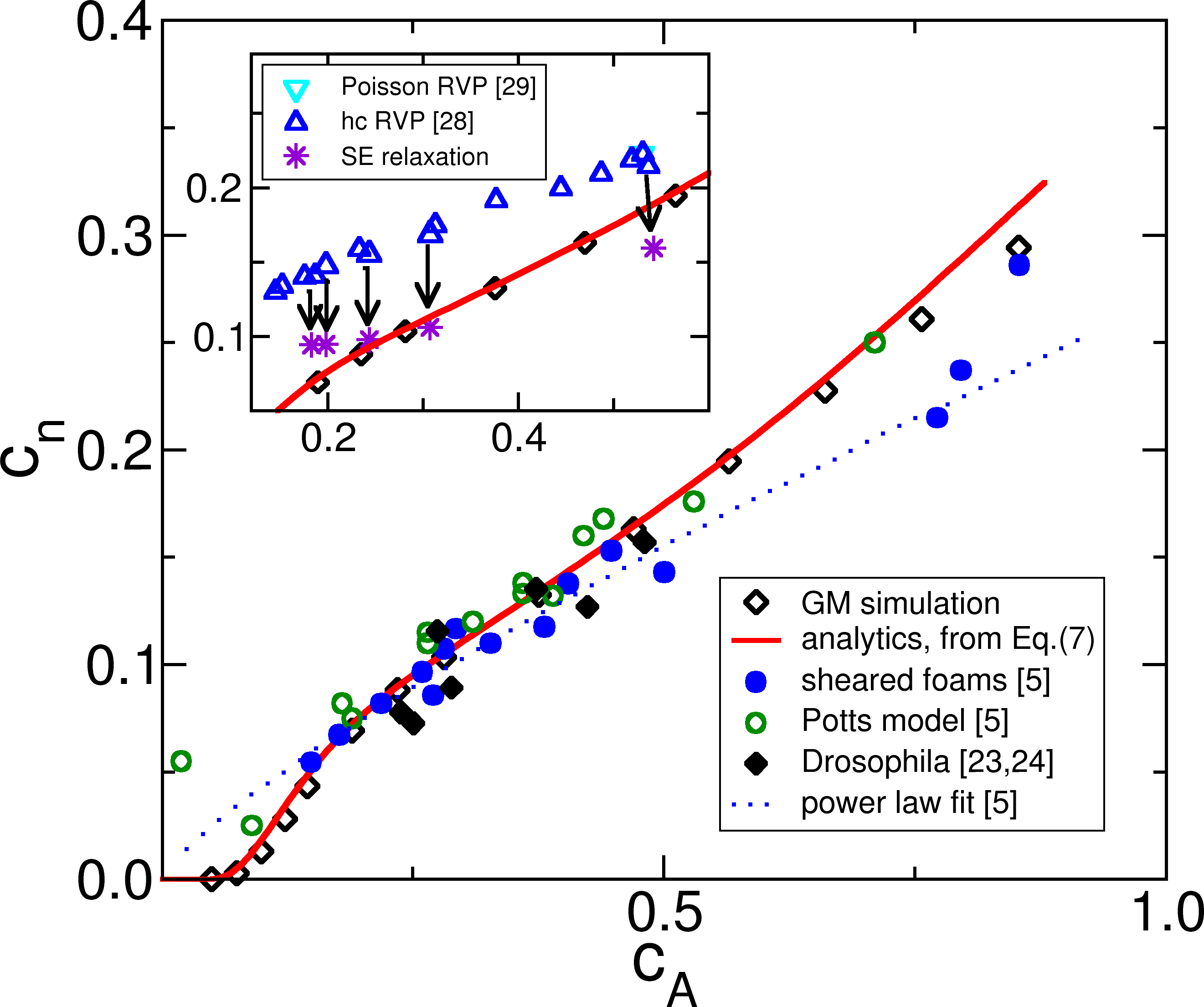}
  \caption{\label{figcnca} Dependence of topological disorder $c_n$ on size disorder $c_A$ in random cellular networks. The analytical theory (solid line) and GM simulations (open diamonds) are in excellent agreement with each other and empirical data (closed symbols: experiments, open symbols: simulations). The inset show that RVP and  hard-core (hc) RVP tilings do not obey this correlation (triangles), but when these structures undergo energy minimization in Surface Evolver (arrows), they approach the GM prediction (stars). The dotted line is the power-law fit of \cite{quilliet08}.}
\end{figure}

 Of particular interest has been the correlation between the number of neighbors $n$ of a domain, a discrete toplogical property, and its area (in 2D) or volume (in 3D), a continuous geometric property. The correlation is intuitive -- larger domains have more neighbors -- but for 2D systems empirical laws have been formulated with a claim to some degree of universality. Lewis' law \cite{lewis28,lewis31} postulates a linear dependence of the average area $\bar{A}_n$ of $n$-sided cells on $n$ for certain systems, while for others nonlinear analogs have been observed \cite{quilliet08,glazier87,lemaitre93}. For many 2D systems with unimodal area distributions, a universal correlation is observed between the coefficient of variation  of the area $c_A$ and that of the neighbor distribution $c_n$, cf.\  \cite{quilliet08} and Fig.~\ref{figcnca}. Fundamental questions have remained unanswered about such laws, in particular (i) whether there is a way to understand them analytically, (ii) whether they are valid for any tiling of the plane, or dependent on physical characteristics of the system, and (iii) whether they can be understood {\em locally} by considering a single domain neighborhood (in the spirit of a mean-field model), or whether spatial correlations are important. We show here that all of these relations follow from a simple, purely local model, that analytical results can be obtained, and that empirical data can be classified according to whether these laws are obeyed.


\begin{figure}[htb]
 \centering
\includegraphics[width=2in]{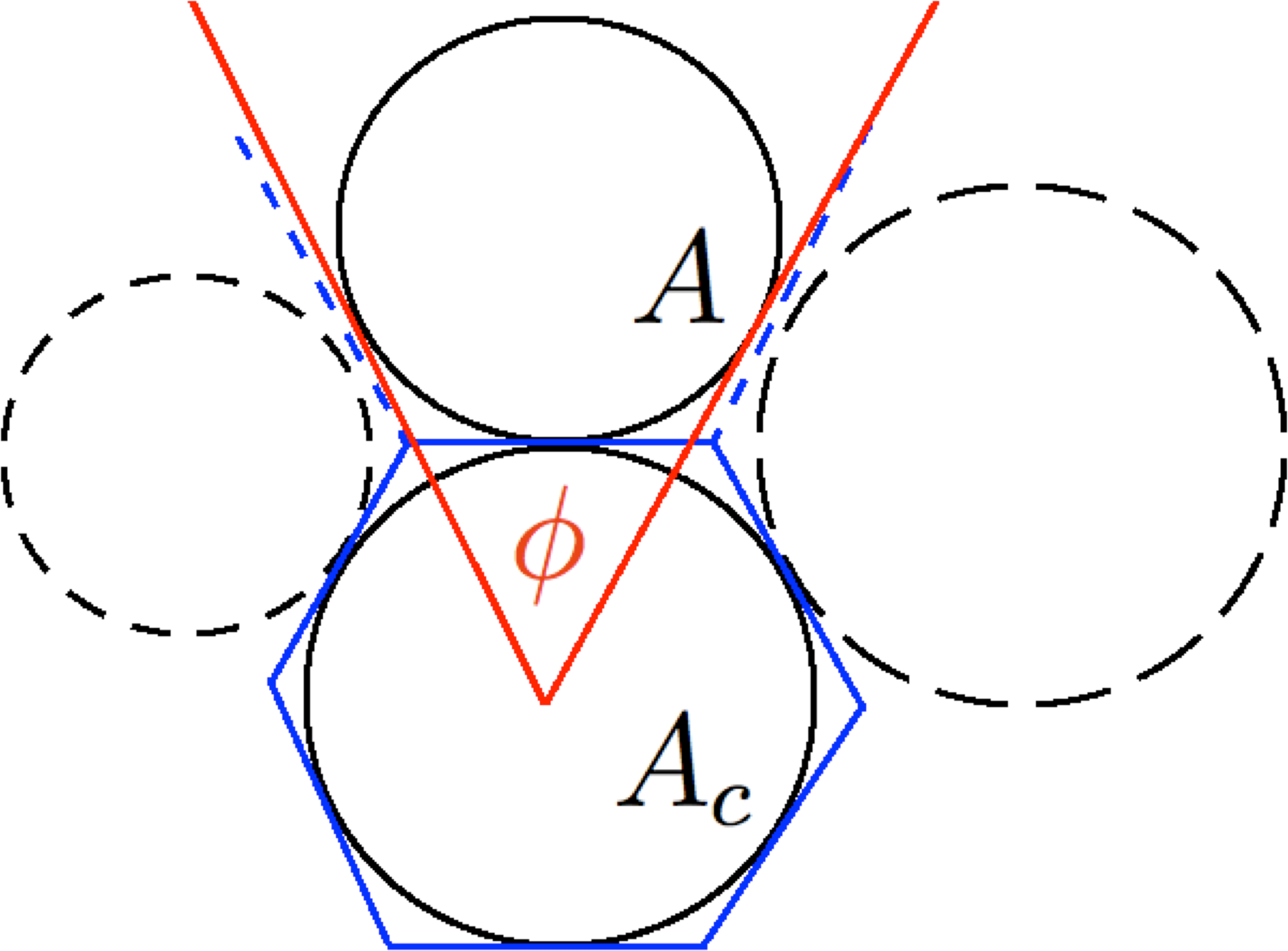}
  \caption{\label{figgm} Local disc packing in the modified 2D granocentric model: a disc of area $A_c$ is surrounded by discs of area $A$ subtending angles $\phi$. The discs serve as templates for polygonal cell construction.}
\end{figure}

Very recently, a local model for 3D systems has been developed incorporating a simple connection between size and neighbor topology. The {\em granocentric model} \cite{clusel09,corwin10} (GM) of Bruji\'c {\em et al.} analyzes the solid angles taken up by spheres neighboring a central sphere and, given the size distribution of spheres, numerically predicts probabilities of numbers of neighbors $n$ and touching neighbors $z$. To self-consistently reproduce the observed $n$, the model must modify the total available solid angle around a sphere ($4\pi$) to $4\pi-\epsilon_3$, where empirically $\epsilon_3\approx 0.32\pi $\cite{corwin10}.
We extend the GM to the 2D case and show that, under certain general approximations, its formalism becomes analytically solvable without free parameters.

%
%
%
%
%
%
%
%
%
%

We consider 2D polydisperse hard discs with area
probability distribution $P(A)$, normalized to the mean $\bar{A}=1$.
As any packing can be converted to a polygonal tiling by constructing cells around the discs (Fig.~\ref{figgm}), this also describes a tiling with a distribution $P_p(A)$. With our main focus on such tilings, we will not model touching neighbors, but only the probability $P_n$ of having $n$ general neighbors. 
A central disc of area $A_c$ is then surrounded by $n$ discs of area $A$ 
(Fig.~\ref{figgm}), where every disc subtends an angle
\be
\phi=g(A)=2\arcsin\left(1/(1+\sqrt{A_c/A})\right)
\label{phidef}
\ee
around the central disc, resulting in the probability distribution of angles
$
f_c(\phi)=P(A(\phi)) (dg^{-1}/d\phi) .
$
Following \cite{corwin10}, we can now compute the conditional probability  $P(n|A_c)$ and the neighbor probability $P_n$ through
\begin{eqnarray}
 P(n|A_c) & = & \int_0^{\phi_{max}} R_{c,n}(\phi)F(\phi_{max}-\phi)d\phi\,,\nonumber\\ 
 P_n & = & \int P(n|A_c)P(A_c)dA_c\,.
 \label{pnacpn}
 \end{eqnarray}
 Here, $R_{c,n} (\phi)$ is the probability of the sum of  $n$ angles from (\ref{phidef}) reaching $\phi$ and $F(\phi_{max}-\phi)\equiv\int_{\phi_{max}-\phi}^\infty f_c(\phi)d\phi$ expresses the probability of the $(n+1)^{th}$ angle exceeding $\phi_{max}$. The maximum angle $\phi_{max}$  available for neighboring discs is again $\not = 2\pi$ because of steric effects. But as Euler's theorem \cite{grunbaum87} demands $\bar{n}=6$ in 2D (for generic three-fold coordinated edges between neighbors), $\phi_{max}=2\pi+\epsilon$ can be determined analytically: To leading order in $c_A\ll 1$, after fitting 5 (touching) neighbors around a disc, there will be a $50\%$ chance of being able to fit a sixth neighbor, i.e., for the unmodified $\phi_{max}=2\pi$ we obtain $\bar{n}=5.5$. To fit, on average, 0.5 additional neighbors, we need to supply the additional average angle subtended by 1/2 disc, i.e., $\epsilon=\pi/6$ and $\phi_{max}=13\pi/6$. Higher-order terms in $c_A$ can be obtained, but we will show that this argument is sufficient in the entire range of relevant $c_A$. 

Eq.~(\ref{pnacpn}) can be evaluated numerically for realistic size and angle distributions. In cellular materials, these are usually gamma or shifted-gamma distributions \cite{kumar05b,lemaitre93,jorjadze11};  Weibull \cite{wyn08b} and other unimodal distributions have also been used. We employ a gamma distribution fit to $f_c(\phi)$  to evaluate (\ref{pnacpn}) numerically; the results are indicated by the label "GM simulations" in subsequent plots. We stress that we have also tried Weibull and other fits with the same mean and variance, and obtained almost indistinguishable results \cite{supmat}.

Beyond numerical evaluation, however, we can derive {\em analytical} results by instead choosing a {\em normal} distribution as fit to $f_c$, again preserving
the first and second moments. The quality of this approximation can be assessed rigorously through an Edgeworth expansion \cite{wal58} and confirmed numerically. As detailed in the Supplemental Material, the approximation is found to give very accurate results {\em independent} of the exact distribution function. We stress that our analytics derive from the single parameter $c_A$ only, with
$c_A\aleq 0.5$ for the vast majority of systems encountered experimentally or theoretically and $c_A\aleq 0.85$ for all instances we could find; thus, expansions in $c_A$ as a small parameter are possible. Expanding $P(A)$ around $\bar{A}=1$, the coefficient of variation of $f_c(\phi)$ is obtained to ${\cal O}(c_A^2)$  from (\ref{phidef}) as
\be
c_{\phi}=c_A/\left([g^{-1}]^\prime[g(1)]g(1)\right)\,.
\label{cphi}
\ee
With Gaussian $f_c$, $R_{c,n}$ is Gaussian as well, with $n$-fold mean and variance, while $F$ is an error function. Then, $P(n|A_c)$  
in (\ref{pnacpn}) 
is obtained using $\int_{-\infty}^{\infty} \exp(-(Ax+B)^2){\rm erf}(Cx+D)dx = (\sqrt{\pi}/A){\rm erf}((AD-BC)/\sqrt{A^2+C^2})$ \cite{sim98b}, yielding $P(n|A_c)=\Phi_{n+1}(c_A,A_c) - \Phi_n(c_A, A_c)$, where
\be
\Phi_n(c_A,A_c)={1\over 2} {\rm erf} \left({n\bar{\phi}(A_c) - \phi_{max} \over \sqrt{2n} \sigma_\phi(c_A,A_c)}\right)\,,
\ee
with $\bar{\phi}(A_c) =2\arcsin(1/(1+\sqrt{A_c}))$ and,  to leading order in $c_A$, the variance
 $\sigma^2_{\phi}=c_A^2s /((1 + s)^2 (2 + s))$, with $s\equiv\sqrt{A_c}$.


Linearizing the argument of this error function around its root in $A_c$, and again making use of a consistent Gaussian approximation with width $c_A$ for $P(A_c)$, we obtain a second integral of the same type, yielding, after further expansion for small $c_A$, the explicit prediction for neighbor probability $P_n=\Psi_{n+1}(c_A)-\Psi_n(c_A)$, with 
\be
\Psi_n(c_A)\!=\!{1\over 2} {\rm erf}\left({\sqrt{2n}\left(1-(2-c_A^2/8)\Sigma\right) \over c_A\!\!\left(\left(1 -\Sigma\right)^2\!\!+\!n(1-c_A^2/8) \Sigma^2\right)^{{1\over 2}}}\right)\!,
\label{psi}
\ee
where $\Sigma\equiv\sin(13\pi/12n)=\sin(\phi_{max}/2n)$ \cite{supmat}.

\begin{figure}[htb]
 \centering
\includegraphics[width=3in]{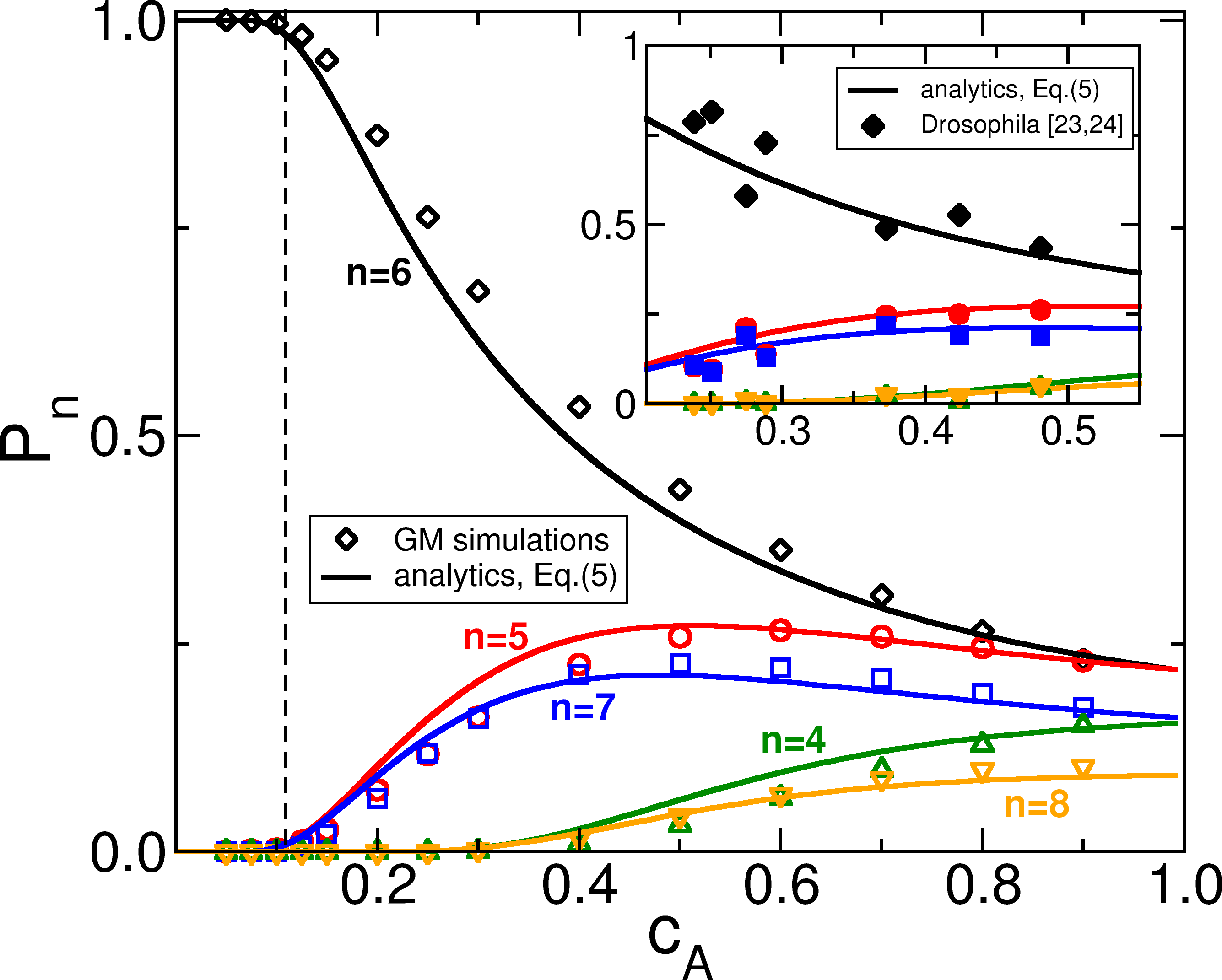}
  \caption{\label{figpn} Neighbor probabilities $P_4$ (triangles up), $P_5$ (circles),  $P_6$ (diamonds), $P_7$ (squares), $P_8$ (triangles down) of discs in the GM simulations. Solid lines are analytical approximations. The plateau of $P_6\approx 1$ quantifies the range of crystallization near monodispersity with $c_{A,crit}=\gamma/\sqrt{3} $ (dashed vertical line). The inset illustrates that the disc $P_n$ are good predictors for numbers of cellular neighbors in random tilings, here for {\em Drosophila} wing tissue cells \cite{classen05,gibson06}.}
\end{figure}

We first interpret these results for discs (without constructing polygonal domains). Plotting the predicted $P_n(c_A)$, we observe a pronounced plateau of hexagonal order in the near-monodisperse case of small $c_A$ (Fig.~\ref{figpn}), so that $P_6\approx 1$ for $c_A\leq c_{A,crit}$. This effect of {\em crystallization}  is well known in simulations of hard-disc packings \cite{ito96,don04,san00}, where a certain amount of polydispersity \cite{sad97,san01} or bidispersity \cite{don06} is necessary to achieve a random structure.  The critical $c_A$ (terminal polydispersity), below which crystallization into ordered domains occurs, has been empirically determined to be $c_{A,crit}\aleq 0.1$ \cite{ito96,sad97,san01}. In the limit $c_A\to 0$, (\ref{psi}) can be further expanded around $n=13/2$, which yields, to excellent accuracy, a compact formula for the fraction of hexagonally coordinated discs,
\be
 P_6 = {\rm erf} \left(\gamma/c_A\right)\,,
\label{p6}
\ee
where $\gamma^2\equiv 2\pi^2/585$. Figure~\ref{figpn} shows that the approximations (\ref{psi}),(\ref{p6}) capture both the numerical results (GM) and the empirically observed feature of crystallization (Eq.~(\ref{p6}) is indistinguishable from (\ref{psi}) with $n=6$). The point of largest curvature of $P_6(c_A)$, where the crystallization plateau begins, is the critical polydispersity $c_{A,crit}=\gamma/\sqrt{3}\approx 0.106$, in very good agreement with empirical values of terminal polydispersity.


\begin{figure}[htb]
 \centering
\includegraphics[width=3in]{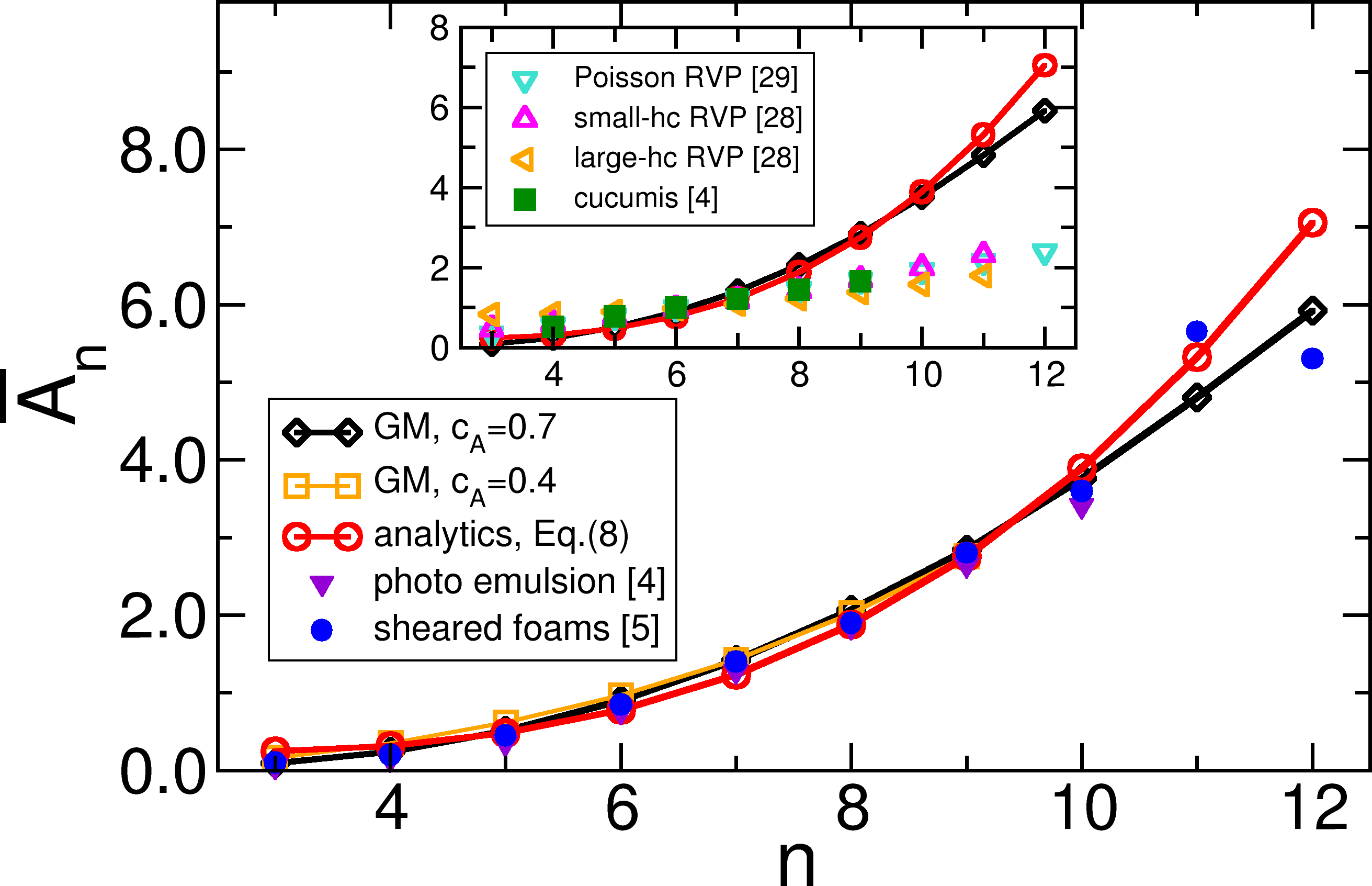}
  \caption{\label{figlewis} The dependence of $\bar{A}_n$ on\ $n$ is nonlinear in disc and most cellular systems. The analytical prediction for this nonlinear Lewis' law agrees well with simulations and experimental data up to $n\approx 10$. Random Voronoi polyhedra  with \cite{zhu01} or without hard cores \cite{brakke05} do not follow this prediction (inset). Lewis' data on cucumber tissue \cite{lewis31} also appears to be in this class.}
\end{figure}

The crystallization threshold divides the regime of nearly-monodisperse objects (where additional orientational disorder, absent from the model, could maintain randomness, e.g.\ in monodisperse cellular tilings) from that of polydisperse objects (where size disorder overwhelms positional disorder \cite{corwin10}). In the latter regime, polygonal cells constructed around the discs should be characterized by the same relations as the discs themselves. As an example, the predicted $P_n$ match experimental data  \cite{classen05,miklius11} for neighbor probabilities in cellular tissues very well, see inset of Fig.~\ref{figpn} (note that topological asymmetry such as $P_5\not = P_7$ is obtained without skewness information from $P(A)$).
In the following, we compare the results of the GM and analytical calculations to empirical size-toplogy correlation data from cellular systems. In order to translate the disc areas into polygonal areas, we simply assume that a disc of area $A$ with $n$ neighbors is inscribed into a regular $n$-gon (with area $A_p=(n/\pi)\tan(\pi/n)A$), and re-normalize the resulting areas. All area-related quantities in Figs.~\ref{figcnca} and \ref{figlewis} thus refer to polygonal areas. While the assumption of regular polygons misses some of the disorder in the system, for cellular systems with domain energy the difference of domain areas from those of regular polygons is typically very small (cf.\ e.g.\ \cite{gra01}).





The expansion that yields (\ref{p6}) also gives the variance $\mu_{2,n}$ of $P_n$,  and thus $c_n=\mu_{2,n}^{1/2}/\bar{n}$. We find
\be
\mu_{2,n}=\sum_{k=1}^\infty (2k-1) {\rm erfc}((2k-1){\gamma\over 2c_A})\,,
\label{mu2n}
\ee
which converges rapidly (even for $c_A\to 1$, truncating the series after the third term is very accurate). Figure~\ref{figcnca} shows the resulting analytical $c_n(c_A)$, together with numerical results and various empirical data points. Also shown is the power-law fit of  \cite{quilliet08}. We see that, for $c_A>c_{A,crit}$, the agreement is excellent for a large variety of systems.
 However, simulations of random Voronoi polygon (RVP) tilings \cite{brakke05,zhu01} significantly disagree with the GM theory and appear to lie on a separate line in $c_n-c_A$ space (inset of Fig.~\ref{figcnca}). 

We suggest that this is caused by the RVP domains' lack of a compact shape, which in turn is due to the absence of an interfacial energy functional. Almost all physical cellular systems, by contrast, exhibit an energy penalty for domain boundaries. We tested this hypothesis by digitizing some of the RVP structures from \cite{zhu01}, using them as initial conditions for an energy minimization of the structure (under constant domain areas) using Surface Evolver \cite{brakke92} (SE) with a uniform interfacial energy density. As the inset of Fig.~\ref{figcnca} shows, the topological disorder $c_n$ indeed diminishes during this process (due to T1 transitions \cite{weaire00}) until the systems approach the observed general $c_n-c_A$ correlation.

The Lewis law correlation can be derived within the GM by computing $\bar{A}_n=\int A_cP(n|A_c)P(A_c) dA_c / P_n$, using the same algorithm as above, including expansions in small $c_A$ and around $n=13/2$ \cite{supmat}. Interestingly, the $c_A$-dependences cancel to leading order, giving the $c_A$-independent approximation
\be
\bar{A}_n=\exp\left({2n(1-\Sigma)\Sigma\over (1-\Sigma)^2+n\Sigma^2}\right)\,.
\label{nllewis}
\ee
This is in agreement with our numerical results that show $\bar{A}_n$ to be insensitive to $c_A$, as well as experimental and other empirical data \cite{quilliet08,lewis31} that confirm this distinctly nonlinear form of Lewis' law (\ref{nllewis}), see Fig.~\ref{figlewis}. The systems for which the $c_n-c_A$ correlation fails (Fig.~\ref{figcnca}) display a linear Lewis' law, clearly disagreeing with the GM theory (inset of Fig.~\ref{figlewis}) and again suggesting two distinct classes of systems distinguishable by compactness of the individual domains. Surface Evolver relaxation of the RVP structures cannot restore the nonlinear Lewis' law, because larger areas are not part of the area distribution, which is not changed by SE.


\begin{figure}[h]
 \centering
\includegraphics[width=2.7in]{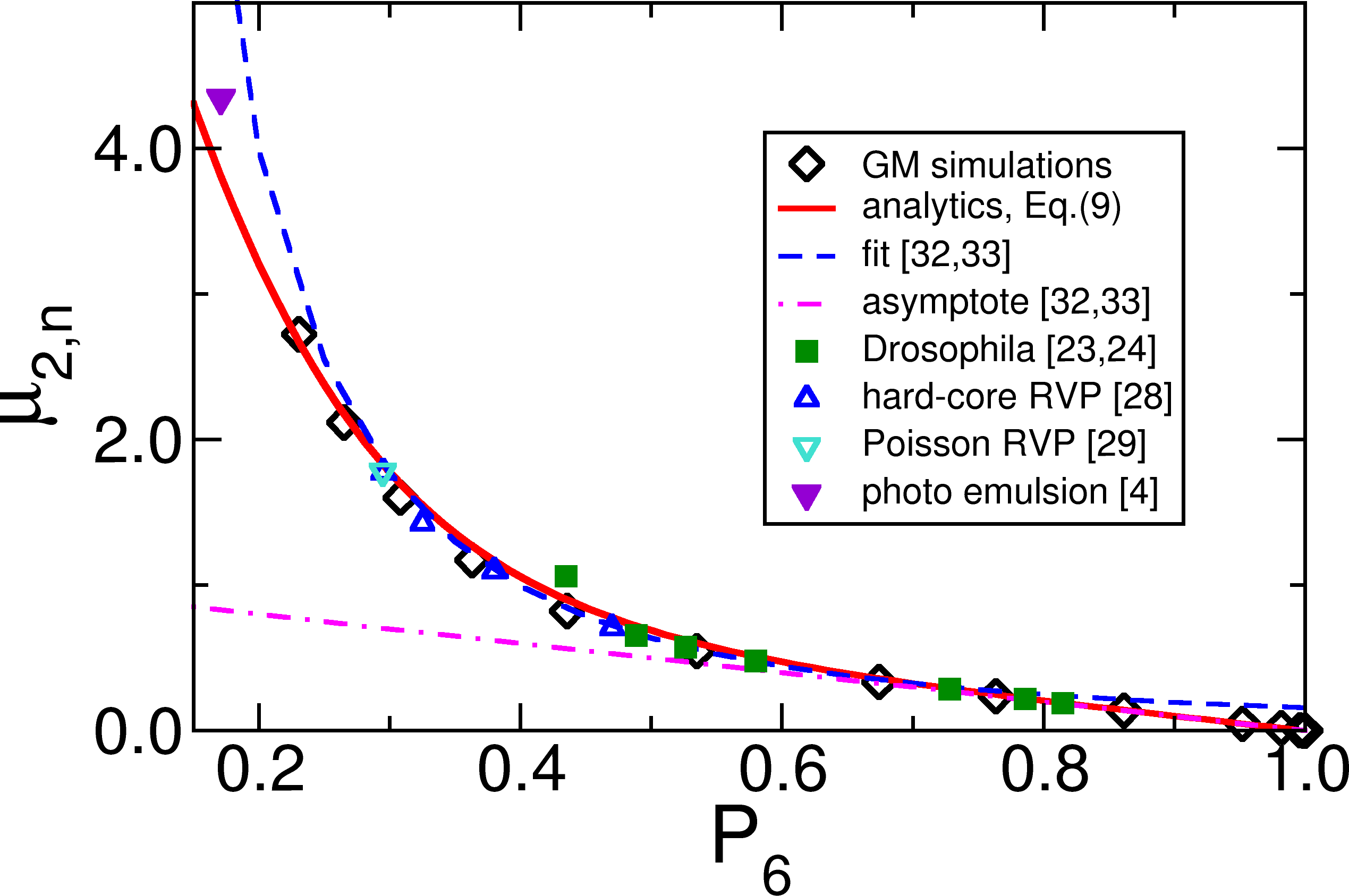}
  \caption{Lema\^itre's law, a robust correlation between $P_6$ and the toplogical variance $\mu_{2,n}$. The analytical theory from the first three terms of (\ref{lemaitre}) is given as solid line, confirmed to great accuracy by many empirical data points and GM simulations (symbols). }
  \label{figlemaitre}
\end{figure}

A relation between $P_6$ and $\mu_{2,n}$ has been reported as Lema\^itre's law \cite{lemaitre91,lemaitre93}, with an empirical two-part fit suggested by \cite{lecaer93,rivier95}. The present formalism again provides concise formulae: from  (\ref{p6}), we obtain the leading-order inversion   $\gamma/c_A\approx \sqrt{\pi} P_6$, which combined with (\ref{mu2n}) yields
\be
\mu_{2,n}(P_6)=\sum_{k=1}^{\infty}(2k-1) {\rm erfc}\left((2k-1){\sqrt{\pi}\over 2} P_6\right)\,.
\label{lemaitre}
\ee
As $P_6\to 1$, this immediately reduces to $\mu_{2,n}= 1-P_6$, in agreement with \cite{lecaer93,rivier95}. As Fig.~\ref{figlemaitre} shows, (\ref{lemaitre}) is a uniformly valid approximation that captures both empirical fits in the reported range of $P_6\ageq 0.3$ \cite{lecaer93,rivier95}. Even though (\ref{lemaitre}) relates two topological quantities, it could only be derived through the area distribution. Lema\^itre's law holds very robustly, being valid for all experimental, simulation, and analytical data. In this sense, all of the analytical information about topological disorder in these systems is contained in $P_6$. 


We have shown that a purely local formalism, a generalization of the granocentric model in two dimensions, yields accurate and general predictions for size-topology correlations in cellular matter, using only the coefficient of variation of the unimodal area distribution. 
Further generalizations are feasible in future work, including neighbor correlations (Aboav-Weaire law) or higher moments of the distributions (e.g.\ to obtain $C_k$ measures \cite{kumar05}). As a leading-order theory, the present formalism does surprisingly well and puts a number of empirical relations on a firm footing.

{\em Acknowledgments:} The authors thank J.\ Bruji\'c and E.\  Vanden-Eijnden for a frank exchange of ideas and results, and invaluable advice and clarifications about the GM.


\begin{thebibliography}{34}%
\makeatletter
\providecommand \@ifxundefined [1]{%
 \@ifx{#1\undefined}
}%
\providecommand \@ifnum [1]{%
 \ifnum #1\expandafter \@firstoftwo
 \else \expandafter \@secondoftwo
 \fi
}%
\providecommand \@ifx [1]{%
 \ifx #1\expandafter \@firstoftwo
 \else \expandafter \@secondoftwo
 \fi
}%
\providecommand \natexlab [1]{#1}%
\providecommand \enquote  [1]{``#1''}%
\providecommand \bibnamefont  [1]{#1}%
\providecommand \bibfnamefont [1]{#1}%
\providecommand \citenamefont [1]{#1}%
\providecommand \href@noop [0]{\@secondoftwo}%
\providecommand \href [0]{\begingroup \@sanitize@url \@href}%
\providecommand \@href[1]{\@@startlink{#1}\@@href}%
\providecommand \@@href[1]{\endgroup#1\@@endlink}%
\providecommand \@sanitize@url [0]{\catcode `\\12\catcode `\$12\catcode
  `\&12\catcode `\#12\catcode `\^12\catcode `\_12\catcode `\%12\relax}%
\providecommand \@@startlink[1]{}%
\providecommand \@@endlink[0]{}%
\providecommand \url  [0]{\begingroup\@sanitize@url \@url }%
\providecommand \@url [1]{\endgroup\@href {#1}{\urlprefix }}%
\providecommand \urlprefix  [0]{URL }%
\providecommand \Eprint [0]{\href }%
\providecommand \doibase [0]{http://dx.doi.org/}%
\providecommand \selectlanguage [0]{\@gobble}%
\providecommand \bibinfo  [0]{\@secondoftwo}%
\providecommand \bibfield  [0]{\@secondoftwo}%
\providecommand \translation [1]{[#1]}%
\providecommand \BibitemOpen [0]{}%
\providecommand \bibitemStop [0]{}%
\providecommand \bibitemNoStop [0]{.\EOS\space}%
\providecommand \EOS [0]{\spacefactor3000\relax}%
\providecommand \BibitemShut  [1]{\csname bibitem#1\endcsname}%
\let\auto@bib@innerbib\@empty
\bibitem [{\citenamefont {Thompson}(1942)}]{thompson42}%
  \BibitemOpen
  \bibfield  {author} {\bibinfo {author} {\bibfnamefont {D.}~\bibnamefont
  {Thompson}},\ }\href@noop {} {\emph {\bibinfo {title} {On Growth and Form}}}\
  (\bibinfo  {publisher} {Cambridge University Press},\ \bibinfo {year}
  {1942})\BibitemShut {NoStop}%
\bibitem [{\citenamefont {Lecuit}\ and\ \citenamefont
  {Lenne}(2007)}]{lecuit07}%
  \BibitemOpen
  \bibfield  {author} {\bibinfo {author} {\bibfnamefont {T.}~\bibnamefont
  {Lecuit}}\ and\ \bibinfo {author} {\bibfnamefont {P.~F.}\ \bibnamefont
  {Lenne}},\ }\href@noop {} {\bibfield  {journal} {\bibinfo  {journal} {Nature
  Reviews Molecular Cell Biologyl}\ }\textbf {\bibinfo {volume} {8}},\ \bibinfo
  {pages} {633} (\bibinfo {year} {2007})}\BibitemShut {NoStop}%
\bibitem [{\citenamefont {Lewis}(1928)}]{lewis28}%
  \BibitemOpen
  \bibfield  {author} {\bibinfo {author} {\bibfnamefont {F.}~\bibnamefont
  {Lewis}},\ }\href@noop {} {\bibfield  {journal} {\bibinfo  {journal} {The
  Anatomical Record}\ }\textbf {\bibinfo {volume} {38}},\ \bibinfo {pages}
  {341} (\bibinfo {year} {1928})}\BibitemShut {NoStop}%
\bibitem [{\citenamefont {Lewis}(1931)}]{lewis31}%
  \BibitemOpen
  \bibfield  {author} {\bibinfo {author} {\bibfnamefont {F.}~\bibnamefont
  {Lewis}},\ }\href {\doibase 10.1002/ar.1090500303} {\bibfield  {journal}
  {\bibinfo  {journal} {The Anatomical Record}\ }\textbf {\bibinfo {volume}
  {50}},\ \bibinfo {pages} {235} (\bibinfo {year} {1931})}\BibitemShut
  {NoStop}%
\bibitem [{\citenamefont {Quilliet}\ \emph {et~al.}(2008)\citenamefont
  {Quilliet}, \citenamefont {Talebi}, \citenamefont {Rabaud}, \citenamefont
  {J.~K\"afer},\ and\ \citenamefont {Graner}}]{quilliet08}%
  \BibitemOpen
  \bibfield  {author} {\bibinfo {author} {\bibfnamefont {C.}~\bibnamefont
  {Quilliet}}, \bibinfo {author} {\bibfnamefont {S.~A.}\ \bibnamefont
  {Talebi}}, \bibinfo {author} {\bibfnamefont {D.}~\bibnamefont {Rabaud}},
  \bibinfo {author} {\bibfnamefont {S.~C.}\ \bibnamefont {J.~K\"afer}}, \ and\
  \bibinfo {author} {\bibfnamefont {F.}~\bibnamefont {Graner}},\ }\href@noop {}
  {\bibfield  {journal} {\bibinfo  {journal} {Phil. Mag. Lett.}\
  }\textbf {\bibinfo {volume} {88}},\ \bibinfo {pages} {651} (\bibinfo {year}
  {2008})}\BibitemShut {NoStop}%
\bibitem [{\citenamefont {Glazier}\ \emph {et~al.}(1987)\citenamefont
  {Glazier}, \citenamefont {Gross},\ and\ \citenamefont {Stavans}}]{glazier87}%
  \BibitemOpen
  \bibfield  {author} {\bibinfo {author} {\bibfnamefont {J.~A.}\ \bibnamefont
  {Glazier}}, \bibinfo {author} {\bibfnamefont {S.~P.}\ \bibnamefont {Gross}},
  \ and\ \bibinfo {author} {\bibfnamefont {J.}~\bibnamefont {Stavans}},\ }\href
  {\doibase 10.1103/PhysRevA.36.306} {\bibfield  {journal} {\bibinfo  {journal}
  {Phys. Rev. A}\ }\textbf {\bibinfo {volume} {36}},\ \bibinfo {pages} {306}
  (\bibinfo {year} {1987})}\BibitemShut {NoStop}%
\bibitem [{\citenamefont {Lema\^itre}\ \emph {et~al.}(1993)\citenamefont
  {Lema\^itre}, \citenamefont {Gervois}, \citenamefont {Troadec}, \citenamefont
  {Rivier}, \citenamefont {Ammi}, \citenamefont {Oger},\ and\ \citenamefont
  {Bideau}}]{lemaitre93}%
  \BibitemOpen
  \bibfield  {author} {\bibinfo {author} {\bibfnamefont {J.}~\bibnamefont
  {Lema\^itre}}, \bibinfo {author} {\bibfnamefont {A.}~\bibnamefont {Gervois}},
  \bibinfo {author} {\bibfnamefont {J.~P.}\ \bibnamefont {Troadec}}, \bibinfo
  {author} {\bibfnamefont {N.}~\bibnamefont {Rivier}}, \bibinfo {author}
  {\bibfnamefont {M.}~\bibnamefont {Ammi}}, \bibinfo {author} {\bibfnamefont
  {L.}~\bibnamefont {Oger}}, \ and\ \bibinfo {author} {\bibfnamefont
  {D.}~\bibnamefont {Bideau}},\ }\href@noop {} {\bibfield  {journal} {\bibinfo
  {journal} {Philosophical Magazine Part B}\ }\textbf {\bibinfo {volume}
  {67}},\ \bibinfo {pages} {347} (\bibinfo {year} {1993})}\BibitemShut
  {NoStop}%
\bibitem [{\citenamefont {Clusel}\ \emph {et~al.}()\citenamefont {Clusel},
  \citenamefont {Corwin}, \citenamefont {Siemens},\ and\ \citenamefont
  {Brujic}}]{clusel09}%
  \BibitemOpen
  \bibfield  {author} {\bibinfo {author} {\bibfnamefont {M.}~\bibnamefont
  {Clusel}}, \bibinfo {author} {\bibfnamefont {E.~I.}\ \bibnamefont {Corwin}},
  \bibinfo {author} {\bibfnamefont {A.~O.~N.}\ \bibnamefont {Siemens}}, \ and\
  \bibinfo {author} {\bibfnamefont {J.}~\bibnamefont {Brujic}},\ }\href@noop {}
  {\bibfield  {journal} {\bibinfo  {journal} {Nature}\ }\textbf {\bibinfo
  {volume} {460}},\ \bibinfo {pages} {611}}\BibitemShut {NoStop}%
\bibitem [{\citenamefont {Corwin}\ \emph {et~al.}(2010)\citenamefont {Corwin},
  \citenamefont {Clusel}, \citenamefont {Siemens},\ and\ \citenamefont
  {Brujic}}]{corwin10}%
  \BibitemOpen
  \bibfield  {author} {\bibinfo {author} {\bibfnamefont {E.~I.}\ \bibnamefont
  {Corwin}}, \bibinfo {author} {\bibfnamefont {M.}~\bibnamefont {Clusel}},
  \bibinfo {author} {\bibfnamefont {A.~O.~N.}\ \bibnamefont {Siemens}}, \ and\
  \bibinfo {author} {\bibfnamefont {J.}~\bibnamefont {Brujic}},\ }\href
  {\doibase 10.1039/C000984A} {\bibfield  {journal} {\bibinfo  {journal} {Soft
  Matter}\ }\textbf {\bibinfo {volume} {6}},\ \bibinfo {pages} {2949} (\bibinfo
  {year} {2010})}\BibitemShut {NoStop}%
\bibitem [{\citenamefont {Grunbaum}\ and\ \citenamefont
  {Shephard}(1987)}]{grunbaum87}%
  \BibitemOpen
  \bibfield  {author} {\bibinfo {author} {\bibfnamefont {B.}~\bibnamefont
  {Grunbaum}}\ and\ \bibinfo {author} {\bibfnamefont {G.~C.}\ \bibnamefont
  {Shephard}},\ }\href@noop {} {\emph {\bibinfo {title} {Tilings and
  Patterns}}}\ (\bibinfo  {publisher} {W.~H.~Freeman and Company},\ \bibinfo
  {address} {New York},\ \bibinfo {year} {1987})\BibitemShut {NoStop}%
\bibitem [{\citenamefont {Kumar}\ and\ \citenamefont
  {Kumaran}(2005{\natexlab{a}})}]{kumar05b}%
  \BibitemOpen
  \bibfield  {author} {\bibinfo {author} {\bibfnamefont {V.~S.}\ \bibnamefont
  {Kumar}}\ and\ \bibinfo {author} {\bibfnamefont {V.}~\bibnamefont
  {Kumaran}},\ }\href {http://dx.doi.org/doi/10.1063/1.2011390} {\bibfield
  {journal} {\bibinfo  {journal} {J. Chem. Phys.}\ }\textbf {\bibinfo {volume}
  {123}},\ \bibinfo {pages} {114501} (\bibinfo {year}
  {2005}{\natexlab{a}})}\BibitemShut {NoStop}%
\bibitem [{\citenamefont {Jorjadze}\ \emph {et~al.}(2011)\citenamefont
  {Jorjadze}, \citenamefont {Pontani}, \citenamefont {Newhall},\ and\
  \citenamefont {Bruji\'c}}]{jorjadze11}%
  \BibitemOpen
  \bibfield  {author} {\bibinfo {author} {\bibfnamefont {I.}~\bibnamefont
  {Jorjadze}}, \bibinfo {author} {\bibfnamefont {L.-L.}\ \bibnamefont
  {Pontani}}, \bibinfo {author} {\bibfnamefont {K.~A.}\ \bibnamefont
  {Newhall}}, \ and\ \bibinfo {author} {\bibfnamefont {J.}~\bibnamefont
  {Bruji\'c}},\ }\href {\doibase 10.1073/pnas.1017716108} {\bibfield  {journal}
  {\bibinfo  {journal} {Proceedings of the National Academy of Sciences}\
  }\textbf {\bibinfo {volume} {108}},\ \bibinfo {pages} {4286} (\bibinfo {year}
  {2011})}\BibitemShut {NoStop}%
\bibitem [{\citenamefont {{Wyn, A.}}\ \emph {et~al.}(2008)\citenamefont {{Wyn,
  A.}}, \citenamefont {{Davies, I. T.}},\ and\ \citenamefont {{Cox, S.
  J.}}}]{wyn08b}%
  \BibitemOpen
  \bibfield  {author} {\bibinfo {author} {\bibnamefont {{Wyn, A.}}}, \bibinfo
  {author} {\bibnamefont {{Davies, I. T.}}}, \ and\ \bibinfo {author}
  {\bibnamefont {{Cox, S. J.}}},\ }\href
  {http://dx.doi.org/10.1140/epje/i2007-10286-0} {\bibfield  {journal}
  {\bibinfo  {journal} {Eur. Phys. J. E}\ }\textbf {\bibinfo {volume} {26}}
  (\bibinfo {year} {2008})}\BibitemShut {NoStop}%
\bibitem [{sup()}]{supmat}%
  \BibitemOpen
  \href@noop {} {}\bibinfo {note} {See the Supplemental Material for more
  details.}\BibitemShut {Stop}%
\bibitem [{\citenamefont {Wallace}(1958)}]{wal58}%
  \BibitemOpen
  \bibfield  {author} {\bibinfo {author} {\bibfnamefont {D.~L.}\ \bibnamefont
  {Wallace}},\ }\href {http://www.jstor.org/stable/2237255} {\bibfield
  {journal} {\bibinfo  {journal} {The Annals of Mathematical Statistics}\
  }\textbf {\bibinfo {volume} {29}},\ \bibinfo {pages} {pp. 635} (\bibinfo
  {year} {1958})}\BibitemShut {NoStop}%
\bibitem [{\citenamefont {Simon}\ and\ \citenamefont
  {Divsalar}(1998)}]{sim98b}%
  \BibitemOpen
  \bibfield  {author} {\bibinfo {author} {\bibfnamefont {M.~K.}\ \bibnamefont
  {Simon}}\ and\ \bibinfo {author} {\bibfnamefont {D.}~\bibnamefont
  {Divsalar}},\ }\href@noop {} {\bibfield  {journal} {\bibinfo  {journal} {IEEE
  Trans. Comm.}\ }\textbf {\bibinfo {volume} {46}},\ \bibinfo {pages} {200}
  (\bibinfo {year} {1998})}\BibitemShut {NoStop}%
\bibitem [{\citenamefont {Ito}(1996)}]{ito96}%
  \BibitemOpen
  \bibfield  {author} {\bibinfo {author} {\bibfnamefont {N.}~\bibnamefont
  {Ito}},\ }\href@noop {} {\bibfield  {journal} {\bibinfo  {journal} {Int. J.
  Mod. Phys. C}\ }\textbf {\bibinfo {volume} {7}},\ \bibinfo {pages} {275}
  (\bibinfo {year} {1996})}\BibitemShut {NoStop}%
\bibitem [{\citenamefont {Donev}\ \emph {et~al.}(2004)\citenamefont {Donev},
  \citenamefont {Torquato}, \citenamefont {Stillinger},\ and\ \citenamefont
  {Connelly}}]{don04}%
  \BibitemOpen
  \bibfield  {author} {\bibinfo {author} {\bibfnamefont {A.}~\bibnamefont
  {Donev}}, \bibinfo {author} {\bibfnamefont {S.}~\bibnamefont {Torquato}},
  \bibinfo {author} {\bibfnamefont {F.~H.}\ \bibnamefont {Stillinger}}, \ and\
  \bibinfo {author} {\bibfnamefont {R.}~\bibnamefont {Connelly}},\ }\href
  {\doibase DOI:10.1063/1.1633647} {\bibfield  {journal} {\bibinfo  {journal}
  {J. Appl. Phys.}\ }\textbf {\bibinfo {volume} {95}},\ \bibinfo {pages} {989}
  (\bibinfo {year} {2004})}\BibitemShut {NoStop}%
\bibitem [{\citenamefont {Santen}\ and\ \citenamefont {Krauth}(2000)}]{san00}%
  \BibitemOpen
  \bibfield  {author} {\bibinfo {author} {\bibfnamefont {L.}~\bibnamefont
  {Santen}}\ and\ \bibinfo {author} {\bibfnamefont {W.}~\bibnamefont
  {Krauth}},\ }\href@noop {} {\bibfield  {journal} {\bibinfo  {journal}
  {Nature}\ }\textbf {\bibinfo {volume} {405}},\ \bibinfo {pages} {550}
  (\bibinfo {year} {2000})}\BibitemShut {NoStop}%
\bibitem [{\citenamefont {Sadr-Lahijany}\ \emph {et~al.}(1997)\citenamefont
  {Sadr-Lahijany}, \citenamefont {Ray},\ and\ \citenamefont {Stanley}}]{sad97}%
  \BibitemOpen
  \bibfield  {author} {\bibinfo {author} {\bibfnamefont {M.~R.}\ \bibnamefont
  {Sadr-Lahijany}}, \bibinfo {author} {\bibfnamefont {P.}~\bibnamefont {Ray}},
  \ and\ \bibinfo {author} {\bibfnamefont {H.~E.}\ \bibnamefont {Stanley}},\
  }\href {\doibase 10.1103/PhysRevLett.79.3206} {\bibfield  {journal} {\bibinfo
   {journal} {Phys. Rev. Lett.}\ }\textbf {\bibinfo {volume} {79}},\ \bibinfo
  {pages} {3206} (\bibinfo {year} {1997})}\BibitemShut {NoStop}%
\bibitem [{\citenamefont {Santen}\ and\ \citenamefont {Krauth}(2001)}]{san01}%
  \BibitemOpen
  \bibfield  {author} {\bibinfo {author} {\bibfnamefont {L.}~\bibnamefont
  {Santen}}\ and\ \bibinfo {author} {\bibfnamefont {W.}~\bibnamefont
  {Krauth}},\ }\href@noop {} {\  (\bibinfo {year} {2001})}\BibitemShut
  {NoStop}%
\bibitem [{\citenamefont {Donev}\ \emph {et~al.}(2006)\citenamefont {Donev},
  \citenamefont {Stillinger},\ and\ \citenamefont {Torquato}}]{don06}%
  \BibitemOpen
  \bibfield  {author} {\bibinfo {author} {\bibfnamefont {A.}~\bibnamefont
  {Donev}}, \bibinfo {author} {\bibfnamefont {F.~H.}\ \bibnamefont
  {Stillinger}}, \ and\ \bibinfo {author} {\bibfnamefont {S.}~\bibnamefont
  {Torquato}},\ }\href {\doibase 10.1103/PhysRevLett.96.225502} {\bibfield
  {journal} {\bibinfo  {journal} {Phys. Rev. Lett.}\ }\textbf {\bibinfo
  {volume} {96}},\ \bibinfo {pages} {225502} (\bibinfo {year}
  {2006})}\BibitemShut {NoStop}%
\bibitem [{\citenamefont {Classen}\ \emph {et~al.}(2005)\citenamefont
  {Classen}, \citenamefont {Anderson}, \citenamefont {Marois},\ and\
  \citenamefont {Eaton}}]{classen05}%
  \BibitemOpen
  \bibfield  {author} {\bibinfo {author} {\bibfnamefont {A.}~\bibnamefont
  {Classen}}, \bibinfo {author} {\bibfnamefont {K.}~\bibnamefont {Anderson}},
  \bibinfo {author} {\bibfnamefont {E.}~\bibnamefont {Marois}}, \ and\ \bibinfo
  {author} {\bibfnamefont {S.}~\bibnamefont {Eaton}},\ }\href@noop {}
  {\bibfield  {journal} {\bibinfo  {journal} {Developmental Cell}\ }\textbf
  {\bibinfo {volume} {9}},\ \bibinfo {pages} {805} (\bibinfo {year}
  {2005})}\BibitemShut {NoStop}%
\bibitem [{\citenamefont {Gibson}\ \emph {et~al.}(2006)\citenamefont {Gibson},
  \citenamefont {Patel}, \citenamefont {Nagpal},\ and\ \citenamefont
  {Perrimon}}]{gibson06}%
  \BibitemOpen
  \bibfield  {author} {\bibinfo {author} {\bibfnamefont {M.}~\bibnamefont
  {Gibson}}, \bibinfo {author} {\bibfnamefont {A.}~\bibnamefont {Patel}},
  \bibinfo {author} {\bibfnamefont {R.}~\bibnamefont {Nagpal}}, \ and\ \bibinfo
  {author} {\bibfnamefont {N.}~\bibnamefont {Perrimon}},\ }\href@noop {}
  {\bibfield  {journal} {\bibinfo  {journal} {Nature}\ }\textbf {\bibinfo
  {volume} {442}},\ \bibinfo {pages} {1038} (\bibinfo {year}
  {2006})}\BibitemShut {NoStop}%
\bibitem [{\citenamefont {{Miklius, M.P.}}\ and\ \citenamefont {{Hilgenfeldt,
  S.}}(2011)}]{miklius11}%
  \BibitemOpen
  \bibfield  {author} {\bibinfo {author} {\bibnamefont {{Miklius, M.P.}}}\ and\
  \bibinfo {author} {\bibnamefont {{Hilgenfeldt, S.}}},\ }\href {\doibase
  10.1140/epje/i2011-11050-7} {\bibfield  {journal} {\bibinfo  {journal} {Eur.
  Phys. J. E}\ }\textbf {\bibinfo {volume} {34}},\ \bibinfo {pages} {50}
  (\bibinfo {year} {2011})}\BibitemShut {NoStop}%
\bibitem [{\citenamefont {Graner}\ \emph {et~al.}(2000)\citenamefont {Graner},
  \citenamefont {Jiang}, \citenamefont {Janiaud},\ and\ \citenamefont
  {Flament}}]{gra01}%
  \BibitemOpen
  \bibfield  {author} {\bibinfo {author} {\bibfnamefont {F.}~\bibnamefont
  {Graner}}, \bibinfo {author} {\bibfnamefont {Y.}~\bibnamefont {Jiang}},
  \bibinfo {author} {\bibfnamefont {E.}~\bibnamefont {Janiaud}}, \ and\
  \bibinfo {author} {\bibfnamefont {C.}~\bibnamefont {Flament}},\ }\href@noop
  {} {\bibfield  {journal} {\bibinfo  {journal} {Phys. Rev. E.}\ }\textbf
  {\bibinfo {volume} {63}},\ \bibinfo {pages} {11402\/1} (\bibinfo {year}
  {2000})}\BibitemShut {NoStop}%
\bibitem [{\citenamefont {Brakke}(2005)}]{brakke05}%
  \BibitemOpen
  \bibfield  {author} {\bibinfo {author} {\bibfnamefont {K.}~\bibnamefont
  {Brakke}},\ }\href@noop {} {\enquote {\bibinfo {title} {200,000,000 random
  voronoi polygons},}\ } (\bibinfo {year} {2005})\BibitemShut {NoStop}%
\bibitem [{\citenamefont {Zhu}(2001)}]{zhu01}%
  \BibitemOpen
  \bibfield  {author} {\bibinfo {author} {\bibfnamefont {H.}~\bibnamefont
  {Zhu}},\ }\href {\doibase doi:10.1080/01418610010032364} {\bibfield
  {journal} {\bibinfo  {journal} {Philosophical Magazine A}\ }\textbf {\bibinfo
  {volume} {81}},\ \bibinfo {pages} {2765} (\bibinfo {year} {1 December
  2001})}\BibitemShut {NoStop}%
\bibitem [{\citenamefont {Brakke}(1992)}]{brakke92}%
  \BibitemOpen
  \bibfield  {author} {\bibinfo {author} {\bibfnamefont {K.}~\bibnamefont
  {Brakke}},\ }\href@noop {} {\bibfield  {journal} {\bibinfo  {journal}
  {Experimental Mathematics}\ }\textbf {\bibinfo {volume} {1}},\ \bibinfo
  {pages} {141} (\bibinfo {year} {1992})}\BibitemShut {NoStop}%
\bibitem [{\citenamefont {Weaire}\ and\ \citenamefont
  {Hutzler}(2000)}]{weaire00}%
  \BibitemOpen
  \bibfield  {author} {\bibinfo {author} {\bibfnamefont {D.}~\bibnamefont
  {Weaire}}\ and\ \bibinfo {author} {\bibfnamefont {S.}~\bibnamefont
  {Hutzler}},\ }\href@noop {} {\emph {\bibinfo {title} {The Physics of
  Foams}}}\ (\bibinfo  {publisher} {Oxford University Press},\ \bibinfo
  {address} {Oxford},\ \bibinfo {year} {2000})\BibitemShut {NoStop}%
\bibitem [{\citenamefont {J.~Lema\^itre}\ and\ \citenamefont
  {Bidcau}(1991)}]{lemaitre91}%
  \BibitemOpen
  \bibfield  {author} {\bibinfo {author} {\bibfnamefont {A.~G.}\ \bibnamefont
  {J.~Lema\^itre}, \bibfnamefont {J.-P.~Troadec}}\ and\ \bibinfo {author}
  {\bibfnamefont {D.}~\bibnamefont {Bidcau}},\ }\href@noop {} {\bibfield
  {journal} {\bibinfo  {journal} {Europhysics Leters}\ }\textbf {\bibinfo
  {volume} {14}},\ \bibinfo {pages} {77} (\bibinfo {year} {1991})}\BibitemShut
  {NoStop}%
\bibitem [{\citenamefont {LeCa\"er}\ and\ \citenamefont
  {Delannay}(1993)}]{lecaer93}%
  \BibitemOpen
  \bibfield  {author} {\bibinfo {author} {\bibfnamefont {G.}~\bibnamefont
  {LeCa\"er}}\ and\ \bibinfo {author} {\bibfnamefont {R.}~\bibnamefont
  {Delannay}},\ }\href@noop {} {\bibfield  {journal} {\bibinfo  {journal} {J.
  Phys. A}\ }\textbf {\bibinfo {volume} {26}},\ \bibinfo {pages} {3931}
  (\bibinfo {year} {1993})}\BibitemShut {NoStop}%
\bibitem [{\citenamefont {Rivier}\ \emph {et~al.}(1995)\citenamefont {Rivier},
  \citenamefont {Schliecker},\ and\ \citenamefont {Dubertret}}]{rivier95}%
  \BibitemOpen
  \bibfield  {author} {\bibinfo {author} {\bibfnamefont {N.}~\bibnamefont
  {Rivier}}, \bibinfo {author} {\bibfnamefont {G.}~\bibnamefont {Schliecker}},
  \ and\ \bibinfo {author} {\bibfnamefont {B.}~\bibnamefont {Dubertret}},\
  }\href@noop {} {\bibfield  {journal} {\bibinfo  {journal} {Acta
  Biotheoretica}\ }\textbf {\bibinfo {volume} {43}},\ \bibinfo {pages} {403}
  (\bibinfo {year} {1995})}\BibitemShut {NoStop}%
\bibitem [{\citenamefont {Kumar}\ and\ \citenamefont
  {Kumaran}(2005{\natexlab{b}})}]{kumar05}%
  \BibitemOpen
  \bibfield  {author} {\bibinfo {author} {\bibfnamefont {V.~S.}\ \bibnamefont
  {Kumar}}\ and\ \bibinfo {author} {\bibfnamefont {V.}~\bibnamefont
  {Kumaran}},\ }\href {\doibase DOI:10.1063/1.2000233} {\bibfield  {journal}
  {\bibinfo  {journal} {J. Chem. Phys.}\ }\textbf {\bibinfo {volume} {123}},\
  \bibinfo {pages} {074502} (\bibinfo {year} {2005}{\natexlab{b}})}\BibitemShut
  {NoStop}%
\end{thebibliography}
\end{document}